\magnification 1200
\centerline {\bf On the Question of Temperature Transformations under Lorentz and 
Galilei Boosts}
\vskip 0.5cm
\centerline {\bf  by Geoffrey L. Sewell}
\vskip 0.5cm
\centerline {\bf Department of Physics, Queen Mary, University of London,}
\vskip 0.2cm
\centerline {\bf  Mile End Road, London E1 4NS, UK}
\vskip 1cm
\centerline {\bf Abstract}
\vskip 0.3cm
We provide a quantum statistical thermodynamical solution of the long standing question 
of  temperature transformations of uniformly moving bodies. Our treatment of this 
question is based on the well established quantum statistical result that the thermal 
equilibrium conditions demanded by both the Zeroth and Second Laws of 
Thermodynamics are precisely those of Kubo, Martin and Schwinger (KMS). We prove 
that, in both the special relativistic and nonrelativistic settings, a state of a body cannot 
satisfy these conditions for different inertial frames with non-zero relative velocity. 
Hence a body that serves as a thermal reservoir, in the sense of the Zeroth Law, in an 
inertial rest frame cannot do so in a laboratory frame relative to which it moves with non-
zero uniform velocity. Consequently, there is no law of temperature transformation under 
either Lorentz or Galilei boosts, and so the concept of temperature stemming from the 
Zeroth Law is restricted to states of bodies in their rest frames.
\vfill\eject
\centerline {\bf 1. Introduction}
\vskip 0.3cm
The question of  how temperature transforms under Lorentz boosts was first addressed by 
Einstein [1] and Planck [2] on the basis of a special relativistic extension of  classical 
thermodynamics. On that basis, they argued that the uniform motion of a body leads to a 
reduction of its observed temperature by the Lorentz contraction factor. In other words, if 
$T_{0}$ is the temperature of a body in its inertial rest frame $K_{0}$, which moves 
with uniform velocity $v$ relative to a laboratory frame, $K_{L}$, then its temperature, 
$T_{L}$, as viewed in $K_{L}$, is given by the formula
$$T_{L}=\bigl(1-v^{2}/c^{2}\bigr)^{1/2}T_{0}.\eqno(1.1)$$
This formula remained unchallenged until, many years later, Ott [3] argued, on the basis 
of a different special relativistic extension of classical thermodynamics, that the 
relationship between $T_{L}$ and $T_{0}$ went the other way, i.e. that
$$T_{L}=\bigl(1-v^{2}/c^{2}\bigr)^{-1/2}T_{0}.\eqno(1.2)$$
Subsequently, Landsberg [4] proposed yet another special relativistic extension of 
thermodynamics, wherein temperature is Lorentz invariant, i.e. $T_{L}=T_{0}$. 
These different approaches to the relativistic extension of thermodynamics led to further 
treatments and comments on the situation by a number of authors, e.g. those of Refs.[5-
7]. In particular, Van Kampen [5] provided a very clear analysis of the assumptions 
underlying the works of Refs. [1-4] and proposed yet another, relativistically covariant 
formulation of classical thermodynamics.
\vskip 0.2cm
At this stage, we remark that no statistical physical considerations were invoked in the 
works [1-7], all of which were based on relativistic extensions of the first and second 
laws of classical thermodynamics. Subsequently, Landsberg and Matsas [8] argued, on 
the basis of a quantum statistical treatment of a particular model, that there is no general 
law of transformation of temperature under Lorentz boosts. Specifically they showed that 
if  $K_{0}$ and $K_{L}$ are two inertial frames as described above, then the coupling 
of  a two-level monopole $M$, at rest in $K_{L}$, to black body radiation, $B$, with 
Planck spectrum in $K_{0}$, drives $M$ into a terminal state that is not one of thermal 
equilibrium for any temperature that varies continuously with the relative velocity $v$. 
Thus, leaving aside the issue of continuity, $B$ behaves as a thermal reservoir only when 
referred to its rest frame. This result is in line with views expressed earlier by Landsberg 
[4]. 
\vskip 0.2cm
The object of the present article is to establish a general version of this result, not only for 
the relativistic case but also for the non-relativistic one, on the basis of a rigorous, model-
independent, quantum statistical treatment of the response of a small test system 
(thermometer!) to its coupling to a moving thermal reservoir. A key ingredient of  this 
treatment is the characterisation of thermal equilibrium states of an arbitrary conservative 
quantum dynamical system by the Kubo-Martin-Schwinger (KMS) condition [9], which 
may formally be expressed, in units where ${\hbar}=k_{Boltzmann}=1$, 
as\footnote*{The mathematically rigorous form of this condition will be specified in 
Section 2.} 
$${\langle}A(t)B{\rangle}={\langle}BA(t+i{\beta}){\rangle},\eqno(1.3)$$
where ${\langle} \ {\rangle}$ denotes expectation value, $A$ and $B$ are arbitrarily 
chosen observables, $A(t)$ is the evolute of $A$ at time $t$ and ${\beta}$ is the inverse 
temperature. This condition serves to extend the definition of canonical equilibrium states 
to infinite systems, which represent natural idealisations of thermal reservoirs, and 
provides a quantum statistical basis of the Zeroth, First and Second Laws of 
Thermodynamics (cf.  [10, Ch. 5] and works cited there). In particular, Kossakowski et al 
[11] proved that the KMS condition is precisely that for which a large (infinite) system, 
${\Sigma}$, behaves as a thermal reservoir, in the sense of the Zeroth Law, i.e. it is the 
necessary and sufficient condition under which ${\Sigma}$ drives any finite test system 
to which it is weakly and transitively\footnote*{The transitivity condition is that the 
coupling serves to provide transitions, whether direct or indirect, between all eigenstates 
of the small system.} coupled into a terminal equilibrium state. 
\vskip 0.2cm
On applying this result to the statistical thermodynamics of moving bodies, we see that if 
${\Sigma}$ is in an equilibrium state in the inertial frame $K_{0}$ and if $S$ is a finite 
test system at rest in another inertial frame $K_{L}$, then ${\Sigma}$ will act as a 
thermal reservoir for $S$, in the sense of the Zeroth Law, if and only if its state satisfies 
the KMS conditions relative to both $K_{0}$ and $K_{L}$, i.e. if the versions of Eq. 
(1.3) relative to these frames are fulfilled for some inverse temperatures ${\beta}_{0}$ 
and ${\beta}_{L}$, respectively. However, we shall prove that this is not possible in 
either the special relativistic or the nonrelativistic setting. Specifically, we shall show 
that, in either setting, {\it a (mixed) state cannot satisfy KMS conditions, whether for the 
same or for different temperatures, relative to two inertial frames whose relative velocity 
is non-zero.} Hence, we conclude that a system that is in an equilibrium state relative to 
$K_{0}$ will not  behave as a thermal state, in the sense of the Zeroth Law, relative to 
$K_{L}$. Thus there is no law of  temperature transformation under either Lorentz or 
Galilei boosts, and so the very concept of temperature is restricted to systems in 
equilibrium in their rest frames. 
\vskip 0.2cm
We present our treatment within the operator algebraic framework of quantum statistical 
physics, pedagogical accounts of which are provided by Refs. [10, 12, 13]. In Section 2, 
we formulate the KMS condition in general operator algebraic terms, specifying there 
both its mathematical and its  thermodynamic significance. In Section 3, we prove our 
main result, namely that a system whose dynamics is either Lorentz or Galilei covariant 
cannot support a state that satisfies the KMS condition relative to different inertial frames 
with non-zero relative velocity. We briefly summarise our conclusions about this result in 
Section 4.
\vskip 0.5cm
\centerline {\bf 2. Preliminaries on Algebraic Structure and the KMS Condition}
\vskip 0.3cm
A conservative quantum dynamical system, ${\Sigma}$, comprises a triple $({\cal 
A},{\alpha},{\cal S})$, where [10,12,13] 
\vskip 0.2cm\noindent
(a) ${\cal A}$ is a $C^{\star}$-algebra whose self-adjoint elements represent its bounded 
observables, 
\vskip 0.2cm\noindent
(b) ${\alpha}$ is a weakly continuous representation of the additive group ${\bf R}$ in 
${\rm Aut}({\cal A})$, the automorphisms of ${\cal A}$, representing the dynamics of 
${\Sigma}$, and
\vskip 0.2cm\noindent
(c) ${\cal S}$ is a folium of  linear, normalised, positive functional on ${\cal A}$, 
representing the  states of ${\Sigma}$. 
\vskip 0.2cm
Thus, the evolute of an observable $A$ at time $t$ is ${\alpha}(t)A$ and the expectation 
value of ${\cal A}$ for the state ${\phi}$ is ${\phi}(A)$, which we also denote by 
${\langle}{\phi};A{\rangle}$. 
\vskip 0.3cm
{\bf Definition 2.1.} A state ${\phi}$ is said to satisfy the KMS condition for inverse 
temperature ${\beta}$ if, for each $A, \ B$ in ${\cal A}$, the function $t({\in}
{\bf R}){\rightarrow}F(t):={\langle}{\phi};B{\alpha}(t)A{\rangle}$ extends to the strip 
${\lbrace}z{\in}{\bf C}{\vert}Im(z){\in}[0,{\beta}]{\rbrace}$, where it is analytic in the 
interior and continuous on the boundaries, and where 
$$F(t+i{\beta})={\langle}{\phi};[{\alpha}(t)A]B{\rangle} \ {\rm and} \ 
F(t)={\langle}{\phi};B{\alpha}(t)A{\rangle} \ {\forall} \ t{\in}
{\bf R}.\eqno(2.1)$$
\vskip 0.2cm
This condition is taken to characterise the thermal equilibrium states of ${\Sigma}$ on 
the following grounds (cf Ref. [10] and works cited therein).
\vskip 0.2cm\noindent
(i) It implies the stationarity of the state ${\phi}$. 
\vskip 0.2cm\noindent
(ii) It corresponds to dynamical and thermodynamical stability conditions that are the 
natural desiderata for equilibrium states. 
\vskip 0.2cm\noindent
(iii) It is just the condition on ${\phi}$ for which an infinite system ${\Sigma}$ behaves 
as a thermal reservoir, in the sense of the Zeroth Law of Thermodynamics.
\vskip 0.3cm
{\bf  KMS Condition and Modular Automorphisms.} An important mathematical 
development in the theory of operator algebras has thrown further light on the KMS 
condition. This development, due originally to Tomita [14] and reformulated by Takesaki 
[15], establishes that, for any faithful normal state ${\psi}$ on a $W^{\star}$-algebra 
${\cal M}$, there is a unique one-parameter group ${\lbrace}{\tau}(t){\vert}t{\in}{\bf 
R}{\rbrace}$ of automorphisms of ${\cal M}$ that satisfy the KMS condition, as given 
by Def. 2.1, with ${\beta}=1$.\footnote*{This result is very simple in the case of a finite 
system where ${\cal M}$ is the algebra of bounded operators in a Hilbert space ${\cal 
H}$. For in this case, a faithful normal state ${\psi}$ corresponds to a density matrix of 
the form ${\rm exp}(-H)$, where $H$ is a lower bounded self-adjoint operator.; and the 
modular automorphisms ${\tau}(t)$ are then implemented by the unitaries ${\rm 
exp}(iHt)$.} These are termed the modular automorphisms for the state ${\psi}$ and are 
characterised by the KMS-type formula 
$${\langle}{\psi};[{\tau}(t)M]N{\rangle}={\langle}{\psi};N{\tau}(t+i)M{\rangle} \ 
{\forall} \ t{\in}{\bf R}, \ M,N{\in}{\cal M}.\eqno(2.2)$$
\vskip 0.2cm
In order to relate the above $C^{\star}$-description of the KMS conditions to these 
automorphisms, we first recall that, by the Gelfand-Neumark-Segal (GNS) construction, 
a state ${\phi}$ on the $C^{\star}$-algebra ${\cal A}$ induces a representation 
${\pi}_{\phi}$ of ${\cal A}$ in a Hilbert space ${\cal H}_{\phi}$ with cyclic vector 
${\Omega}_{\phi}$. In particular, if ${\phi}$ is stationary, the automorphisms 
${\alpha}$ are implemented by a unitary representation $U_{\phi}$ of ${\bf R}$ in 
${\cal H}_{\phi}$, as defined by the formula
$$U_{\phi}(t){\pi}_{\phi}(A){\Omega}_{\phi}=[{\pi}_{\phi}({\alpha}(t)A)]
{\Omega}_{\phi} \ {\forall} \ t{\in}{\bf R}, \ A{\in}{\cal A}.\eqno(2.3)$$
Further, the state ${\phi}$ has a  canonical extension ${\hat {\phi}}$ to the $W^{\star}$-
algebra ${\pi}_{\phi}({\cal A})^{{\prime}{\prime}}$, defined by the formula 
$${\hat {\phi}}(R)=({\Omega}_{\phi},R{\Omega}_{\phi}) \ {\forall} \ 
R{\in}{\pi}_{\phi}({\cal A})^{{\prime}{\prime}}.\eqno(2.4)$$ 
Moreover, in the case where ${\phi}$ is stationary, the dynamical automorphism group 
${\alpha}$ has a canonical extension ${\alpha}_{\phi}$ to 
${\pi}_{\phi}({\cal A})^{{\prime}{\prime}}$, that is implemented by the unitaries 
$U_{\phi}$, i.e.
$${\alpha}_{\phi}(t)R=U_{\phi}(t)RU_{\phi}(-t) \ {\forall} \ t{\in}{\bf R}, \ 
R{\in}{\pi}_{\phi}({\cal A})^{{\prime}{\prime}}\eqno(2.5)$$
and, in particlular,
$${\alpha}_{\phi}(t)\bigl[{\pi}_{\phi}(A)\bigr]={\pi}_{\phi}\bigl({\alpha}(t)A\bigr) \ 
{\forall} \ A{\in}{\cal A}, \ t{\in}{\bf R}.\eqno(2.6)$$
Thus, by Def. 2.1 and Eqs. (2.4)-(2.6), if ${\phi}$ satisfies the KMS condition with 
respect to the automorphism group ${\alpha}$ at inverse temperature ${\beta}$, then 
${\hat {\phi}}$ fulfills the corresponding KMS condition with respect to the 
automorphisms ${\alpha}_{\phi}$ of ${\pi}_{\phi}({\cal A})^{{\prime}{\prime}}$, i.e.
$${\langle}{\hat {\phi}};[{\alpha}_{\phi}(t)R]S{\rangle}=
{\langle}{\hat {\phi}};S{\alpha}_{\phi}(t+i{\beta})R{\rangle} \ {\forall} \ t{\in}{\bf R}, 
\ R,S{\in}{\pi}_{\phi}({\cal A})^{{\prime}{\prime}}.\eqno(2.7)$$ 
Moreover, it has also been established [9], on the basis of the KMS condition on 
${\phi}$, that the vector ${\Omega}_{\phi}$ is cyclic not only for the algebra 
${\pi}_{\phi}({\cal A})^{{\prime}{\prime}}$ but also for its commutant 
${\pi}_{\phi}({\cal A})^{\prime}$. Thus it is both cyclic and separating for 
${\pi}_{\phi}({\cal A})^{{\prime}{\prime}}$, which signifies that ${\hat {\phi}}$ is a 
faithful normal state on the latter algebra. Consequently, it follows from a comparison of 
Eqs. (2.2) and (2.7), with ${\cal M}={\pi}_{\phi}({\cal A})^{{\prime}{\prime}}$ and 
${\psi}={\hat {\phi}}$, that the dynamical group ${\alpha}_{\phi}$ is the time-rescaled 
version of the modular group ${\tau}$ given by the formula
$${\alpha}_{\phi}(t/{\beta})={\tau}(t) \ {\forall} \ t{\in}{\bf R}.\eqno(2.8)$$
\vskip 0.5cm
\centerline {\bf 3. Statistical Thermodynamics of Moving Bodies}
\vskip 0.3cm
We take our generic model of a reservoir to be an infinitely extended quantum system, 
${\Sigma}$, whose properties we now treat in both the special relativistic and the non-
relativistic settings. 
\vskip 0.3cm
\centerline {\bf 3.1. The Special Relativistic Model.}
\vskip 0.3cm 
We assume that the relativistic system ${\Sigma}$ occupies the Minkowski space-time 
$X$. Employing units for which $c=1$, we denote the space-time coordinates of a point 
$x$ of $X$, relative to an inertial frame $K$, by  
${\lbrace}x^{\mu}{\vert}{\mu}=0,1,2,3{\rbrace}$, with $x^{0}=t$. Thus, defining $u$ 
to be the unit vector along the time direction, i.e.
$$u=(1,0,0,0),\eqno(3.1)$$
the time coordinate relative to $K$ is
$$t=x.u:=x^{\mu}u_{\mu}.\eqno(3.2)$$
We define the transformations $T(a)$ and $L(v)$ of $X$ corresponding to
space-time translation by $a$ and velocity boost by $v \ ({\in}(-1,1))$
along $Ox^{1}$, respectively, by the formulae
$$T(a)x=x+a\eqno(3.3)$$
and  
$$L(v)x=\bigl({{x^{0}-vx^{1}}\over (1-v^{2})^{1/2}},
{{x^{1}-vx^{0}}\over (1-v^{2})^{1/2}},x^{2},x^{3}\bigr).\eqno(3.4)$$ 
It follows from these formulae that 
$$L(v)T(a)L(-v)=T(L(v)a) \ {\forall} \ a{\in}X, \ v{\in}(-1,1).\eqno(3.5)$$
In particular, by Eqs. (3.1), (3.4) and (3.5), the Lorentz transform of the time translation 
$T(tu)$ is given by the formula
$$L(v)T(tu)L(-v)=T(tu^{\prime}),\eqno(3.6)$$
where
$$u^{\prime}=L(v)u=(1-v^{2})^{-1/2}(1,-v,0,0),\eqno(3.7)$$
which is just the unit time vector in the reference frame $K^{\prime}$ that moves with 
velocity $v$ along $Ox^{1}$ relative to $K$. We shall henceforth assume that 
$v{\neq}0$.
\vskip 0.2cm
We employ the scheme of Haag and Kastler [16] for the operator algebraic formulation of 
the model of  ${\Sigma}$. Thus, we start by defining ${\cal L}$ to be the set of all 
bounded, open subsets of $X$, and we assign to each such subset ${\Lambda}$ a 
$C^{\star}$-algebra ${\cal A}_{\Lambda}$, the algebra of bounded observables for the 
space-time region ${\Lambda}$. We assume that this satisfies the following two 
canonical requirements of isotony and local commutativity (Einstein causality!), i.e. that
\vskip 0.2cm\noindent
(a) ${\cal A}_{\Lambda}$ is isotonic with respect to ${\Lambda}$, i.e. ${\cal 
A}_{\Lambda}{\subset}{\cal A}_{{\Lambda}^{\prime}}$ if 
${\Lambda}{\subset}{\Lambda}^{\prime}$; and
\vskip 0.2cm\noindent
(b) if ${\Lambda}$ and ${\Lambda}^{\prime}$ have spacelike separation, then the 
elements of ${\cal A}_{\Lambda}$ and ${\cal A}_{{\Lambda}^{\prime}}$ 
intercommute.
\vskip 0.2cm
In view of the isotony property (a), the union, ${\cal A}_{\cal L}$, of the local algebras 
${\cal A}_{\Lambda}$ is a normed $^{\star}$-algebra, whose norm completion is a 
$C^{\star}$-algebra ${\cal A}$. We take ${\cal A}$ to be the algebra of the quasi-local 
bounded observables of ${\Sigma}$. We then define the state space ${\cal S}$ to 
comprise the positive, normalised, linear functionals on ${\cal A}$. We assume that the 
dynamics of the system is Poincare covariant and thus that the Poincare group ${\cal P}$ 
is represented by automorphisms of ${\cal A}$. In particular, space-time translations and 
Lorentz boosts are represented by homomorphisms ${\xi}$ and ${\lambda}$ of the 
additive groups $X$ and ${\bf R}$,  respectively, in ${\rm Aut}({\cal A})$ that satisfy 
the canonical analogue of Eq. (3.5), namely
$${\lambda}(v){\xi}(a){\lambda}(-v)={\xi}\bigl(L(v)a\bigr) \ {\forall} \ a{\in}X, \ 
v{\in}{\bf R}.\eqno(3.8)$$
The subgroup of ${\xi}(X)$ corresponding to time translations relative to $K$ is 
${\alpha}({\bf R})$, with
$${\alpha}(t)={\xi}(tu) \ {\forall} \ t{\in}{\bf R}\eqno(3.9)$$
and $u$ is the unit time vector defined by Eq. (3.1). It follows immediately from Eqs. 
(3.7)-(3.9) that
$${\lambda}(v){\alpha}(t){\lambda}(-v)={\alpha}^{\prime}(t):={\xi}(tu^{\prime}) \ 
{\forall} \ t{\in}{\bf R}, \ v{\in}{\bf R},\eqno(3.10)$$ 
which signifies that ${\alpha}^{\prime}(t)$ is the Lorentz transform of ${\alpha}(t)$ 
corresponding to the velocity boost $v$ and is therefore the automorphism representing 
time translation by $t$ in the reference frame $K^{\prime}$.
\vskip 0.2cm
By Def. 2.1 and Eq. (3.9), the KMS condition on a state ${\phi}$, relative to the frame 
$K$ at inverse temperature ${\beta}$, is that
$${\langle}{\phi};[{\xi}(tu)A]B{\rangle}=
{\langle}{\phi};B{\xi}\bigl((t+i{\beta})u\bigr){\rangle} \ {\forall} \ t{\in}{\bf R}, \ 
A,B{\in}{\cal A}.\eqno(3.11)$$ 
Correspondingly, the KMS condition on ${\phi}$, relative to the frame $K^{\prime}$ at 
inverse temperature ${\beta}^{\prime}$, is that
$${\langle}{\phi};[{\xi}(tu^{\prime})A]B{\rangle}=
{\langle}{\phi};B{\xi}\bigl((t+i{\beta}^{\prime})u^{\prime}\bigr){\rangle} \ {\forall} \ 
t{\in}{\bf R}, \ A,B{\in}{\cal A}.\eqno(3.12)$$ 
\vskip 0.3cm
{\bf Definition 3.1.}We say that space-time translations act {\it non-trivially} in the GNS 
representation ${\pi}_{\phi}$ of this state if, for any non-zero element $a$ of $X$, there 
exists some $A{\in}{\cal A}$ and $s{\in}{\bf R}$ for which 
${\pi}_{\phi}\bigl({\xi}(sa)A\bigr){\neq}{\pi}_{\phi}(A)$.  
\vskip 0.3cm
{\bf Proposition 3.1.} {\it Let ${\phi}$ be a state on ${\cal A}$ that satisfies the KMS 
condition relative to the frame $K$ at inverse temperature ${\beta}$. Then, assuming that 
space-time  translations act non-trivially in the GNS representation of  ${\phi}$, there is 
no inverse temperature ${\beta}^{\prime}$ for which this state satisfies the KMS 
condition relative to the frame $K^{\prime}$.} 
\vskip 0.3cm
{\bf Proof.} Assume that ${\phi}$ satisfies both of  the KMS conditions (3.11) and 
(3.12). Then it follows from Eq. (2.8) that the canonical extensions ${\alpha}_{\phi}({\bf 
R})$ and ${\alpha}_{\phi}^{\prime}({\bf R})$ of ${\alpha}({\bf R})$ and 
${\alpha}^{\prime}({\bf R})$, respectively, to ${\pi}_{\phi}({\cal 
A})^{{\prime}{\prime}}$ are related to the (unique) modular automorphism group 
${\tau}({\bf R})$ for ${\phi}$ by the formula
$${\alpha}_{\phi}(t/{\beta})={\alpha}_{\phi}^{\prime}(t/{\beta}^{\prime})={\tau}(t) \ 
{\forall} \ t{\in}{\bf R}.\eqno(3.13)$$
Hence, by Eqs. (2.6), (3.9),  (3.10) and (3.13), 
$${\pi}_{\phi}\bigl({\xi}(tu/{\beta})A\bigr)=
{\pi}_{\phi}\bigl({\xi}(tu^{\prime}/{\beta}^{\prime})A\bigr) \ {\forall} \ 
A{\in}{\cal A}, \ t{\in}{\bf R}.\eqno(3.14)$$
On replacing $t$ by ${\beta}^{\prime}t$ and $A$ by ${\xi}(-tu^{\prime})A$ in this 
equation, we see that
$${\pi}_{\phi}\Bigl({\xi}\bigl(t({\beta}^{-1}{\beta}^{\prime}u-u^{\prime})\bigr)\Bigr)
={\pi}_{\phi}(A) \ {\forall} \ A{\in}{\cal A}, \ t{\in}{\bf R}.\eqno(3.15)$$
Consequently, by the assumption that space-time translations act non-trivially in the GNS 
representation of ${\phi}$, 
$${\beta}^{\prime}u={\beta}u^{\prime}.\eqno(3.16)$$
By Eqs. (3.1) and (3.7), this last equation signifies that
$${\beta}^{\prime}={\beta}(1-v^{2})^{-1/2} \ {\rm and} \ v{\beta}(1-v^{2})^{-1/2}=0.
\eqno(3.17)$$
In view of the finiteness of ${\beta}$ and the condition that ${\vert}v{\vert}<1$, the 
second of these equations cannot be satisfied for non-zero $v$. This completes the proof 
of the Proposition.. 
\vskip 0.3cm
{\bf Comment.} This Proposition, when combined with the result of Ref. [11], 
establishes that, if ${\Sigma}$ is in thermal equilibrium relative to a rest frame $K$, then 
it does not satisfy the demand of the Zeroth Law relative to a moving frame 
$K^{\prime}$. Hence there is no temperature transformation law under Lorentz boosts. 
\vskip 0.5cm
\centerline {\bf  3.2. The Non-Relativistic Model}
\vskip 0.3cm
For the non-relativistic model ${\Sigma}$ (cf. [9, 10, 12, 13]), the space-time is 
$X{\times}{\bf R}$, where now $X$ is a Euclidean space and  the dynamics is Galilei, 
rather than Poincare, covariant. Thus, space-time points are denoted by $(x,t)$, where 
$x({\in}X)$ and $t({\in}{\bf R})$ are their spatial and temporal components, 
respectively. We denote by $S(a), \ T(b)$ and $G(v)$ the transformations of 
$X{\times}{\bf R}$ corresponding to space translation by $a$, time translation by $b$ 
and  vector-valued velocity boost $v$, respectively, as given by the formulae
$$S(a)(x,t)=(x+a,t),\eqno(3.18)$$
$$T(b)(x,t)=(x,t+b)\eqno(3.19)$$
and  
$$G(v)(x,t)=(x-vt,t).\eqno(3.20)$$
Thus $S, \ T$ and $G$ are representations of the additive groups $X, \ {\bf R}$ and $X$, 
respectively. It follows from these formulae that $S(a)$ and $T(b)$ intercommute and 
that
$$G(v)T(b)G(-v)=S(-vb)T(b) \ {\forall} \ v{\in}Y, \ t{\in}{\bf R}.\eqno(3.21)$$
\vskip 0.2cm
The $C^{\star}$-algebra, ${\cal A}$, of quasi-local bounded observables is formulated  
by the prescription of Section 3.1, but now with $X$ a Euclidean space. We assume that 
space translations, time translations\footnote*{We remark that this assumption 
concerning time translations is not always satisfied by non-relativistic models and that, 
more generally, their evolution is of the $W^{\star}$-dynamical kind [17, 18]. However, 
we employ the $C^{\star}$-dynamical description here for simplicity, noting that a 
parallel treatment, which yields the same results, can be carried through on the 
$W^{\star}$ scheme of Ref. [18].} and Galilei boosts are represented by 
homomorphisms ${\sigma}, \ {\alpha}$ and ${\gamma}$ of the additive groups $X, \ 
{\bf R}$ and $X$,  respectively, into ${\rm Aut}({\cal A})$. Thus, assuming that the 
dynamics of the system is Galilei covariant, the intercommutativity of $S(a)$ and $T(b)$ 
implies that of ${\sigma}(x)$ and ${\alpha}(t)$, and the following canonical analogue of 
Eq. (3.21) is satisfied. 
$${\gamma}(v){\alpha}(t){\gamma}(-v)={\alpha}^{\prime}(t):=
{\sigma}(-vt){\alpha}(t)\eqno(3.22)$$
Thus,  ${\alpha}^{\prime}(t)$ is the Galilei transform of ${\alpha}(t)$, corresponding to 
the velocity boost $v$. In other words, as ${\alpha}$ represents the time translations 
relative to a rest frame $K, \ {\alpha}^{\prime}$ represents them relative to an inertial 
frame $K^{\prime}$ that moves with velocity $v$. We shall henceforth assume that 
$v{\neq}0$.
\vskip 0.2cm
We now proceed to the following non-relativistic analogues of Def. 3.1 and Prop. 3.1.
\vskip 0.3cm
{\bf Definition 3.2.}We say that space-time translations act {\it non-trivially} in the GNS 
representation ${\pi}_{\phi}$ of a state ${\phi}$ on ${\cal A}$ if, for any non-zero 
element $(a,b)$ of $X{\times}{\bf R}$, there exists some $A{\in}{\cal A}$ and 
$s{\in}{\bf R}$ for which 
${\pi}_{\phi}\bigl({\sigma}(sa){\alpha}(sb)A\bigr){\neq}{\pi}_{\phi}(A)$.  
\vskip 0.3cm
{\bf Proposition 3.2.} {\it Let ${\phi}$ be a state on ${\cal A}$ that satisfies the KMS 
condition relative to the frame $K$ at inverse temperature ${\beta}$. Then, assuming that 
space-time  translations act non-trivially in the GNS representation of  ${\phi}$, there is 
no inverse temperature ${\beta}^{\prime}$ for which thi state satisfies the KMS 
condition relative to the frame $K^{\prime}$.} 
\vskip 0.3cm
{\bf Proof.} Assume that ${\phi}$ satisfies the KMS conditions relative to $K$ and 
$K^{\prime}$ at inverse temperatures ${\beta}$ and ${\beta}^{\prime}$, respectively. 
Then it follows from Def. 2.1 and Eqs. (2.6), (2.8) and (3.22) that 
$${\pi}_{\phi}\bigl({\alpha}(t/{\beta})A\bigr)={\pi}_{\phi}
\bigl({\sigma}(-vt/{\beta}^{\prime}){\alpha}(t/{\beta}^{\prime})A\bigr)=
{\tau}(t)\bigl({\pi}_{\phi}(A)\bigr) \ {\forall} \ 
A{\in}{\cal A}, \ t{\in}{\bf R},\eqno(3.23)$$
where ${\tau}({\bf R})$ is the modular automorphism group for the canonical extension 
${\hat {\phi}}$ of ${\phi}$ to ${\pi}_{\phi}({\cal A})^{{\prime}{\prime}}$. On 
replacing $A$ by ${\alpha}(-t/{\beta})A$ in this equation, we see that
$${\pi}_{\phi}(A)={\pi}_{\phi}\Bigl({\sigma}(-vt/{\beta}^{\prime})
{\alpha}\bigl(t[({\beta}^{\prime})^{-1}-{\beta}^{-1}]\bigr)A\Bigr) \ {\forall} \ A{\in}
{\cal A}, \ t{\in}{\bf R}.\eqno(3.24)$$
Consequently, by the assumption that space-time translations act non-trivially in the GNS 
representation of ${\phi}$, it follows from this equation and Def. 3.2 that
$$v/{\beta}^{\prime}=0 \ {\rm and} \ {\beta}^{\prime}={\beta}.\eqno(3.25)$$
Since these equations cannot be satisfied under our assumptions that ${\beta}$ is finite 
and $v$ is non-zero, it follows that the state ${\phi}$ cannot satisfy the KMS conditions 
relative to both $K$ and $K^{\prime}$. 
\vskip 0.3cm
{\bf Comment.} This Proposition, when combined with the result of Ref. [11], 
establishes that, if ${\Sigma}$ is in thermal equilibrium relative to a rest frame $K$, then 
it does not satisfy the demand of the Zeroth Law relative to a moving frame 
$K^{\prime}$. Hence there is no temperature transfromation law under Galilean boosts..
\vskip 0.5cm
\centerline {\bf 4. Concluding Remarks}
\vskip 0.3cm
We have provided a general, quantum statistical treatment of the thermodynamics of 
moving bodies, that is based on 
\vskip 0.2cm\noindent
(1) the Poincare or Galilei covariance of the dynamics; 
\vskip 0.2cm\noindent
(2) the characterisation of thermal equilibrium by the KMS condition; and
\vskip 0.2cm\noindent
(3) the simple relationship of that condition to the modular theory of Tomita and 
Takesaki.  
\vskip 0.2cm
On this basis, which underpins the Zeroth Law as well as the First and Second ones, we 
have shown here that there is no temperature transformation law under either Lorentz of 
Galilei boosts. Consequently, in both the special relativistic and the non-relativistic 
settings, the concept of temperature, stemming from the Zeroth Law, is restricted to 
equilibrium states of systems in their rest frames. 
\vskip 1cm
\centerline {\bf References}
\vskip 0.3cm\noindent
[1] A. Einstein: Jahrb. Radioaktivitaet und Elektronik {\bf 4}, 411 (1907)
\vskip 0.2cm\noindent
[2] M. Planck: Ann. Der Physik {\bf 26}, 1 (1908)
\vskip 0.2cm\noindent
[3] H. Ott: Zeits. Phys. {\bf 175}, 70 (1963)
\vskip 0.2cm\noindent
[4] P. T. Landsberg: Nature {\bf 212}, 571 (1966); Nature {\bf 214}, 903 (1967)
\vskip 0.2cm\noindent
[5] N. G. Van Kampen: Phys. Rev. {\bf 173}, 295 (1968)
\vskip 0.2cm\noindent
[6] T. W. B. Kibble: Nuovo Cimento {\bf 41B}, 72 (1966)
\vskip 0.2cm\noindent
[7] H. Callen and G. Horowitz: Amer. J. Phys. {\bf 39}, 938 (1971)
\vskip 0.2cm\noindent
[8] P. T. Landsberg and G. E. A. Matsas: Phys. Lett. A {\bf 223}, 401 (1996)
\vskip 0.2cm\noindent
[9] R. Haag, N. M. Hugenholtz and M. Winnink: Commun. Math. Phys. {\bf 5}, 215 
(1967)
\vskip 0.2cm\noindent
[10] G. L. Sewell: {\it Quantum Mechanics and its Emergent Macrophysics}, Princeton 
Univ. Press, Princeton, 2002.
\vskip 0.2cm\noindent
[11] A. Kossakowski, A. Frigerio, V. Gorini and M. Verri: Commun. Math. Phys. 
{\bf 57}, 97, 1977
\vskip 0.2cm\noindent
[12] G. G. Emch: {\it Algebraic Methods in Statistical mechanics and Quantum Field 
Theory}, Wiley, New York, 1972
\vskip 0.2cm\noindent
[13] W. Thirring: {\it Quantum Mechanics of Large Systems}, Springer, New York, 1983
\vskip 0.2cm\noindent
[14] M. Tomita: {\it Standard Forms of Von Neumann Algebras}, Vth Functional 
Analysis Symposium of the Math. Soc. of Japan, Sendai (1967)
\vskip 0.2cm\noindent
[15] M. Takesaki: {\it Tomita\rq s Theory of Modular Hilbert Algebras and its 
Applications}, Lec. Notes in Maths. {\i28}, Springer, Berlin, Heidelberg, New York, 
1970.
\vskip 0.2cm\noindent
[16] R. Haag and D. Kastler: J. Math. Phys. {\bf 5}, 848 (1964)  
\vskip 0.2cm\noindent
[17] D. A. Dubin and G. L. Sewell: J. Math. Phys. {\bf 11}, 2990 (1970)
\vskip 0.2cm\noindent
[18] G. L. Sewell: Lett. Math. Phys. {\bf 6}, 209 (1982)

\end